\documentstyle[aps,prl,multicol,psfig]{revtex}
\tighten
\begin{document}

%
%
\title{Stripe formation: A quantum critical point for cuprate superconductors}
\author{{C. Castellani, \underline{C. Di Castro}},  and M. Grilli}
\address{Istituto di Fisica della Materia e
Dipartimento di Fisica, Universit\`a di Roma ``La Sapienza'',\\
Piazzale A. Moro 2, 00185 Roma, Italy}
\maketitle
%
%
\begin{abstract} 
We discuss the effects of a quantum critical point located nearby
optimum doping and related to local charge segregation (stripe phase).
The fluctuations in the critical region produce at the same time
a strong pairing mechanism and a non-Fermi liquid behavior in the
normal phase above the superconducting critical temperature.
Superconductivity is a stabilizing mechanism against
charge ordering, i.e. the incommensurate charge density wave
quantum critical point is unstable with respect to superconductivity.
A complete scenario for the cuprates is presented.
\end{abstract}
%
%
\pacs{PACS:74.20.-z, 74.72.-h,71.27.+a}

%
%

\begin{multicols}{2}

\section{INTRODUCTION}
The presence of a critical point at zero temperature (Quantum Critical Point,
QCP) ruling the physical properties of the superconducting cuprates was 
repeatedly suggested \cite{sachdev,pines,varma,CDG,perali,evora}:
In the quantum critical region above the QCP
no energy scale besides temperature controls the physics and 
strong critical fluctuations can mediate singular 
interactions between quasiparticles, providing both a strong pairing 
mechanism and a source of non-Fermi liquid (non-FL) behavior in the
normal-state \cite{CDG}.

A natural candidate where to place
the QCP is the region nearby optimum doping where the
highest critical temperatures and the best non-FL metal
appear. Indeed the existence of a QCP  near optimum doping
is supported by several experimental findings. In particular
recent transport measurements in ${\rm La_{2-x}Sr_{x}CuO_{4}}$ (LSCO)
 investigated the normal phase of these
systems when superconductivity is suppressed
 under strong magnetic field \cite{boebinger}. This analysis 
shows the existence of a QCP near optimum
doping. Evidences in this sense
 are also provided by neutron scattering \cite{aeppli},
by many experiments showing qualitative changes of behavior
taking place near optimum doping between underdoped and overdoped
samples and by the fact that several quantities
(resistivity, Hall number, uniform susceptibility) show
a scaling behavior with a typical energy scale, which 
vanishes at optimum doping \cite{refall}.

Many indications exist that the above QCP involves charge ordering.
 Generically, it seems quite likely that charge degrees of freedom
play a major role, since the disordered region of this QCP
coincides with the highly metallic overdoped regime. Experimentally
a direct observation of charge-driven ordering
was possible by neutron scattering
\cite{tranquada1},
in ${\rm La_{1.48}Nd_{0.4}Sr_{0.12}CuO_4}$ 
where the related Bragg peaks were detected. For this specific
compound the low-temperature-tetragonal lattice structure 
pins the charge-density waves (CDW) and gives static order and 
semiconducting behavior (see also the case of 
${\rm La_{1.88}Ba_{0.12}CuO_4}$).
Increasing the Sr content at fixed Nd concentration, the pinning
effect is reduced leading to metallic and superconducting behavior.
In this latter case, the existence of dynamical incommensurate CDW
(ICDW) fluctuations is suggested by the presence of incommensurate dynamic
spin scattering, although the charge peaks are too weak to be observed. 
In this regard, also
${\rm La_{2-x}Sr_xCuO_4}$ is expected to display dynamical charge
fluctuations with a doping-dependent spatial modulation as
observed in the magnetic scattering \cite{yamada}.
ICDW have also been inferred from extended X-ray absorption
fine structure (EXAFS) experiments both in optimally
doped LSCO \cite{bianconi1} and ${\rm Bi_2Sr_2CaCu_2O_{8+x}}$
(Bi-2212) \cite{bianconi2}.

The above phenomenology supports the scenario of a 
proximity to an ICDW
transition \cite{CDG,perali,evora}, which in the absence of superconductivity
would be located at zero temperature near
the optimum doping. 
 Large superconducting critical temperatures are
most naturally expected where strong attractive critical fluctuations
occur. Then, superconductivity takes place and 
 the CDW instability is hindered. The CDW instability
can only show its effects above $T_c$ as responsible 
in the quantum critical region for the best non-FL
behavior occurring near optimum doping in all classes of cuprates.
The presence of an ICDW-QCP
is not alternative to the existence of 
an AF-QCP and the two QCP's control the behavior of the system
at different dopings.  
It is worth noting that within the ICDW scenario, the
ICDW stripes constitute the substrate to sustain the antiferromagnetic
(AF) fluctuations far away from the AF-QCP.
In the underdoped compounds, 
the ICDW instability would occur at finite temperature,
were it not for the quenching due to superconducting local fluctuations
which can give rise to the 
appearance of charge and spin gaps, as experimentally found 
\cite{marshall,harris,campuzano,rossat,jullien,loram}
at a temperature $T^*$ above $T_c$. According to this proposal,
the underlying 
hindered charge instability provides a strongly temperature dependent
pairing potential \cite{evora}. 
This explains the high crossover temperature $T^*$
of the ($d$-wave)
pair formation and the peculiar doping dependence of $T^*$,
which strongly increases with decreasing doping, 
\cite{marshall,harris,campuzano} while 
the superconducting gap at T=0 remains nearly constant \cite{harris,loram}. 

\section{The ICDW instability: The formation mechanism}

The occurrence of an ICDW-QCP is theoretically substantiated by 
considering the interplay between electronic phase separation (PS)
and long-range Coulombic (LRC) repulsion. 

PS is a common feature of strongly correlated
electron systems: It has been found in magnetic models 
\cite{tj1,tj2,CCCDGR,dagotto},
in models with nearest-neighbour Coulomb 
interactions \cite{weak,coul1,coul2,coul3,ruckenstein,simonaligia},  
 and in models with phononic interactions 
\cite{GC,longr,markiewicz1,markiewicz2}. 
Indeed a strong on-site correlation 
drastically renormalizes the kinetic energy, which would tend to delocalize
the carriers. Then short-range interactions 
(magnetic, phononic, nearest-neighbor coulombic...)
introducing effective attractions between the carriers
may dominate and give rise to charge aggregation in highly doped 
metallic regions together with charge depletion in spatially separated 
(magnetically ordered) regions with no itinerant charges. 
As it was first noticed in \cite{CCCDGR,coul1},  
pair formation always occur before and nearby PS. Therefore
``phase separation
does not spoil superconductivity. Pair formation at fixed doping
is rather a stabilizing mechanism with respect to phase separation 
between the normal phases'' \cite{CCCDGR}. 

It was then pointed out \cite{physc} that 
LR Coulomb forces effectively oppose the separation
of charged particles suppressing long-wavelenght density fluctuations.
This may lead to either dynamical slow density 
fluctuations \cite{physc} or
ICDW  \cite{coul2,CDG,emery}.
In this latter case {\it microscopic} calculations 
in the limit of strong electron-electron interaction were
carried out both in the single-band infinite-U Hubbard model 
in the presence of an Holstein 
electron-phonon coupling (Hubbard-Holstein model) \cite{CDG,longr}
and in the infinite-U three-band Hubbard model
extended with a nearest-neighbor (Cu-O) repulsion V \cite{BBG}.
Similar results
were found in both models thus confirming that the interplay between PS and
LRC forces provide a robust mechanism for
charge ordering instabilities. 
Despite the different dynamics, 
in both cases a similar singular attraction arises of the form
\begin{equation}
\Gamma ({\bf q},\omega) \approx 
\tilde{U} - {1 \over 4} \sum_\alpha \frac{V}{\kappa^2+
\omega_{\bf q}^{\alpha}
- i\gamma \omega} 
\label{fitgamlr}
\end{equation}
where $\tilde{U}$ is the residual repulsive interaction 
between the quasiparticles and $\gamma$ is a damping parameter.
The sum is over the  four equivalent vectors of the CDW 
instability ${\bf q}^{\alpha}=(\pm q_c,0),(0,\pm q_c)$ and 
$\omega_{\bf q}^{\alpha} =
2(2-\cos(q_x-q_x^{\alpha})-\cos(q_y-q_y^{\alpha}))$.
This expression is used to reproduce the behavior $\sim -1/(\kappa^2+
(q_x-q_x^{\alpha})^2+(q_y-q_y^{\alpha})^2)$ for
$q\rightarrow q^{\alpha}$ while mantaining the lattice periodicity. 

The mass term $\kappa^2=a(\delta-\delta_c)$ 
is found  \cite{CDG} to be linear in the doping deviation from the
critical concentration. For reasonable band-structure parameters 
 the instability first occurs at $\delta_c\approx 0.2$
with $q_c \lesssim 1$.
From our analysis the rather large density of states near the $(\pm \pi,0)$
and $(0,\pm \pi)$ points tends to favor instabilities at or close
to the (1,0) or (0,1) directions.
However, due to the rather small values of $q_c$, the scattering is 
quite strong, although non-singular, in all directions
for $\vert {\mbox{\boldmath $q$}} \vert \approx \vert 
{\mbox{\boldmath $q$}}_c \vert$.

\section{The ICDW-QCP in the absence of pairing}
In the light of the generality of the results
of the previous section, we now discuss the properties
of the ICDW-QCP without a specific reference to a microscopic model,
thus pointing out the robust generic features of the scenario.
For the sake of definiteness we will assume the form 
(\ref{fitgamlr}) to be generically valid and we will also assume 
a simple Gaussian behavior of the QCP.

Around a QCP three regions are usually identified
(thin solid lines in Fig. 1).
\begin{figure}
\centerline{\hbox{\psfig{figure=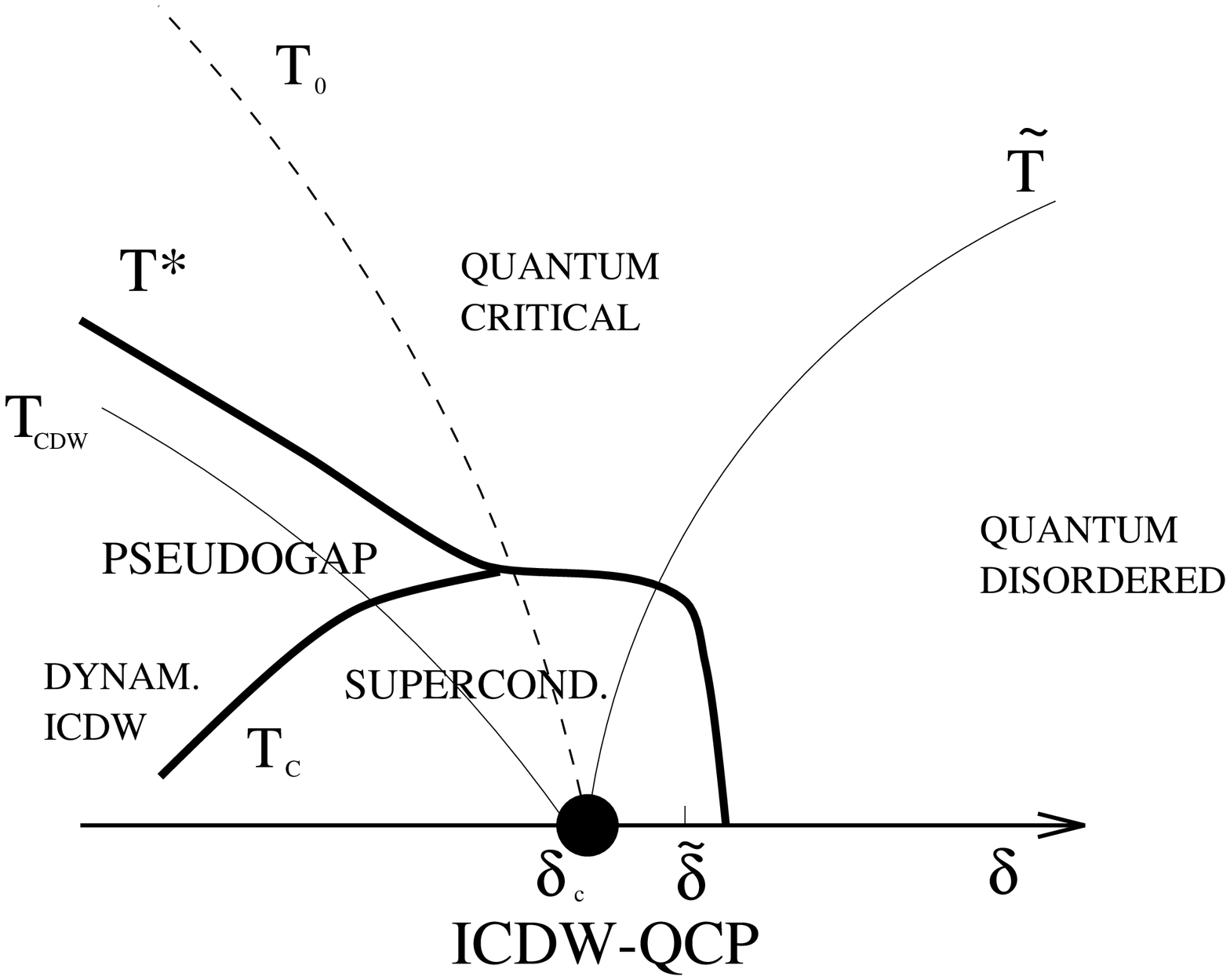,width=7.cm,angle=-90}}}
{\small FIG. 1: Schematic structure of the 
temperature vs. doping $\delta$ phase diagram
around the ICDW-QCP.
On the right, $T<\tilde{T}$: Quantum disordered region
[$\xi^{-2}\approx (\delta-\delta_c)$]; In the middle,
$T>T_{CDW}, \tilde{T}$: 
Quantum critical (classical gaussian) region 
[$\xi^{-2}\approx T$]; On the left, $T<T_{CDW}$: ``ordered'' ICDW phase.
The heavy line indicates the region of local (pseudogap)
or coherent (superconducting) pairing. In the presence of pairing,
i.e. for $T<T^*$, the ICDW ``order'' is purely dynamical.}
\label{FIG1}
\end{figure}
The quantum disordered (QD) regime on the right would in the present case 
correspond to the overdoped region, where 
$\kappa^2 =\xi^{-2} \sim a(\delta-\delta_c)^{2\nu}$. 
Within the adopted 
classical gaussian approximation, we take $\nu=1/2$.
Increasing the temperature one enters in the so-called
classical gaussian regime (quantum critical, QC), where
$\kappa^2$ becomes a function of $T$, 
$\kappa^2 \sim b T^{(d+z-2)/z}$, where $d$ is the spatial dimension and 
$z$ is the dynamical critical index. The proper $z$ is $z=2$ 
for CDW as one sees from  the fluctuation propagator (\ref{fitgamlr}).
$a$ and $b$ are model-dependent positive constants. 
$\tilde{T}$ marks the crossover between the two regimes.

The region on the left of the QCP would generically correspond to
the ordered ICDW phase occurring below a critical temperature
$T_{CDW}(\delta)$ starting from the QCP 
at $\delta_c(T=0)$.
The true critical line is depressed with respect to
its mean-field expression (sketched by the dashed line 
$T_0$ of Fig. 1). The region between the two curves 
$T_0$ and $T_{CDW}$ is dominated by strong precursor effects. 

The quantum disordered region on the right corresponds
to the overdoped region of the cuprates, where a crossover
to a low-temperature FL takes place.
On other hand the classical gaussian region around
optimum doping is characterized by the absence
of energy scales, but the temperature. 
Here the best non-FL behavior is obtained.
In particular, with the scattering of the form 
given in Eq. (\ref{fitgamlr}), a linear-in-T resistivity
is expected in $d=2$, while for $d=3$, $\rho(T)\sim T^{3/2}$.
For magnetically mediated scattering, 
in Ref. \onlinecite{hlubina} the objection was raised
that only few "hot" points would feel strong scattering
contributing to the above behavior. Generically, 
all other points would contribute to the lower 
$T^2$ behavior. However, for ICDW, the fact that 
typical $q_c$ are fairly small and the strong isotropic
character of $\Gamma(q)$ make this objection less relevant.

It is worth noting that the quantum critical behavior in transport
may extend to the overdoped region down to low
temperatures well below $\tilde{T}$. Indeed, whereas the critical
crossover line $\tilde{T}(\delta)$ is established by
matching the behavior of $\kappa^2$
as a function of $(\delta-\delta_c)$ and $T$ in the QD and 
QC regions respectively, transport crucially 
involves temperature, i.e. frequency,
scales. Then, the temperature below which the fluctuations
feel the presence of the mass $\kappa^2$ is determined by the
condition $\gamma \omega \sim \gamma
T \sim \kappa^2$. Clearly for large enough
$\gamma$ quite low temperatures $T_{FL}$ can be reached before
the singular interaction is cut off leading to FL behavior
($\rho \propto T^2$). 

Within our scenario the temperature $T_0$ marks the
onset of the ICDW precursors and is characterized by 
a loss of spectral weight at low energies
giving rise to a uniform decrease of the density of states
near the Fermi energy. This would show up as the well known decrease
of the uniform magnetic susceptibility below a characteristic
temperature, which vanishes by approaching from below the
optimum doping \cite{johnston}.
In underdoped ${\mathrm {YBa_2Cu_3O_{6+x}}}$ (YBCO)
compounds, this last temperature has also been 
put in correspondence \cite{ito,batlogg}
 with the temperature below which
the planar resistivity $\rho_{ab}$ deviates from its linear behavior.

The scenario presented so far should find a wider
physical correspondence whenever the superconducting pair
formation is forbidden. Boebinger {\it et al.}  \cite{boebinger} 
report the results of transport experiments
in LSCO, in the presence of high
magnetic fields, allowing to access the normal phase
underlying the superconducting region. They assess the
presence of a metal-insulator transition in the underdoped 
region ending near optimum doping ($x\approx 0.17$) at $T=0$.
Actually, their work not only shows the existence of 
QCP {\it different from the} AF-QCP, but also 
provides a clear evidence that the  nature of this
instability is related to charge-ordering.

The experimental line separating the planar
metal from the insulator would correspond
in our picture to the ``true'' $T_{CDW}$
critical temperature as a function of doping,
which manifests itself when superconductivity is suppressed.
We also observe that the
resistivity curve for $x=0.12$, i.e. near the "magic" doping 
$1/8$, displays an insulating behavior at a much higher temperature
than for values of $x$ immediately nearby.
This shows that commensurability indeed plays a relevant role
in establishing the insulating phase, thereby indicating that
spatial order of the charge degrees of freedom should be involved
also away from commensurability.
This would also agree with the observation made in Ref. 
\cite{boebinger} that the system is rather clean
($k_F {l} \sim 13$)
and the disorder by itself cannot be the source of the transition
\cite{notadisorder}

Concerning the location of the QCP, we would expect that both the
metal-insulator transition line (which we identified with $T_{CDW}$) 
and the temperature of the onset of pseudogap (identified by $T_0$),
would end to the same point at $T=0$. Fig. 3 of Ref. \cite{boebinger}
seem to assign a value of $x_c=0.17$ to the critical doping $\delta_c$.
On the other hand from Fig. 8 of Ref. \cite{batlogg} reporting 
the temperature $T_0$ for many physical quantities,
one would deduce a somewhat larger value $x_c\gtrsim 0.2$
(in Ref. \cite{batlogg} our $T_0$ is indicated by $T^*$).
However, this discrepancy can be solved by noting that 
the resistivity curve at $x=0.17$ 
in Fig. 1 of Ref. \cite{boebinger} is likely still affected by 
superconductivity at the lowest temperature, 
and the metal-insulator transition
at this filling seems to occur at a still finite temperature
(of about 15 K). Therefore, 
the published data of Fig. 1(b) in Ref. \cite{boebinger}
are compatible with a shift of the 
metal-insulator transition at $T=0$, 
towards dopings that are larger than the assigned value.

An additional observation can be made by contrasting the behavior of
LSCO systems with La-doped ${\rm Bi_2Sr_2CuO_y}$ (Bi-2201) materials
\cite{ando}. These latter systems are slightly overdoped. 
 Therefore it is not surprising that the low temperature behavior 
of the planar resistivity is always metallic in Bi-2201. 
On the other hand they are much more anisotropic than the LSCO systems.
Therefore it is again natural to find that $\rho_c$ is always 
semiconducting in  Bi-2201. This strongly twodimensional
character also accounts for the robust linear behavior of the resisitivity
which is expected in the quantum-critical region above 
a twodimensional QCP.
On the contrary, the transverse hopping, being larger 
in LSCO, easily becomes coherent by increasing
doping, giving rise to threedimensional metallic behavior.
In this case the observed $T^{3/2}$ behavior for the planar resistivity
in the overdoped regime 
is explained as the result of the quantum critical 
behavior in three dimension \cite{moriya}.

\section{The ICDW-QCP in the presence of pairing}
The singular interaction of Eq.(\ref{fitgamlr}), when considered
in the particle-particle channel, provides a strong
pairing mechanism in the proximity of the critical point. 
 By allowing for  superconducting pairing
the thick solid lines in Fig. 1 schematically represent 
the crossover ($T^*$) to local pairing
or the coherent-superconducting transition
($T_c$) in the phase diagram near the ICDW-QCP, thus eliminating
the $T_{CDW}$ and part of the $\tilde{T}$ line.

The most apparent and generic result is that pairing
has $d$-wave symmetry \cite{perali} and, being mediated
by an interaction  
rapidly varying with $\kappa^2$ [cf. Eq.(\ref{fitgamlr})],
strongly depends on temperature or doping.
Roughly, the $d$-wave  becomes favorable since the
average repulsion felt by the $s$-wave paired electrons 
exceeds the loss in condensation energy due to
the vanishing of the order parameter along the nodal regions.
Among the $d$ waves, the $d_{x^2-y^2}$ is preferred because 
the nodes occur in regions with small density of states.

The superconducting critical temperature 
evaluated in the BCS approach \cite{notaBCS}
shows a strong increase  upon decreasing $\kappa^2$.
The actual behavior of $T_c$ is then obtained by
introducing the doping and temperature dependence
of $\kappa^2\equiv \kappa^2(\delta-\delta_c,T)$.
An additional (less important) doping dependence is due to 
the variation of the chemical potential with respect
to the van Hove singularity
(VHS). In the quantum disordered phase the pairing potential 
becomes stronger and stronger  by decreasing doping towards $\delta_c$
and a rapid increase in 
$T_c [\simeq T_c(\kappa^2(\delta-\delta_c,T=0))$]  will result.
At a given doping $\tilde{\delta}\gtrsim \delta_c$ the 
BCS superconducting temperature will reach the
crossover line $\tilde{T}$ separating the quantum disordered from
the classical gaussian region.
In this latter region $\kappa^2$ weakly depends on doping
and a plateau in $T_c$ is reached.

In this scheme the maximal critical temperature is obtained near
 the quantum-disordered/quantum-critical crossover
($\delta_{opt}\approx \tilde{\delta}$).
Of course this is only an estimate depending on
the use of a weak coupling BCS scheme and of a model dependent evaluation
of $\kappa^2 (\delta-\delta_c, T)$.
Nevertheless, we remark that the 
rapid variation of $T_c$ with doping in the overdoped region,
observed in many cuprates, and the plateau
near optimum doping are naturally captured by our description.

Notice that in discussing $T_c$ vs doping we have used the fact
that the main doping dependence is via $\kappa^2$. 
Strong variations of $T_c$ with doping 
 are hardly obtained in terms of a dependence on 
band parameters only (specifically, tuning the VHS). 
They are instead quite natural 
in the context of proximity to an instability, 
where doping controls the 
effective potential itself and not only the density of states.
This agrees with 
the experimental finding that at the maximum $T_c$ 
the VHS is not at the 
Fermi energy but below it \cite{shenreview}.

The region on
the left of the mean-field critical curve for the ICDW
transition is characterized by strong thermal
fluctuations leading to ICDW precursors.
The ICDW fluctuations in the underdoped region
become critical in the proximity of the line $T_{CDW}(\delta)$
where the ICDW transition would occur in the absence of
superconducting pairing. 
Approaching $T_{CDW}$ the attractive fluctuations would lead to the 
formation of (local) pairs at the
curve $T^*(\delta)$. As a consequence of strongly
paired quasiparticles, below $T^*(\delta)$
pseudogap effects will show up 
as seen in many experiments (NMR, ARPES, 
optical conductivity, specific
heat, ...). However, despite
the strong pairing, the true superconducting critical
temperature is lower than $T^*$ and it
decreases inside the underdoped psudogap region.
This occurrence is schematically depicted in Fig. 3 by the
bifurcation of the heavy line. 

Experimentally it is observed \cite{loram,harris,campuzano}
that the low-temperature gap in the underdoped cuprates
weakly depends on doping, while $T^*$ increases fast by
decreasing $\delta$. 
 Although a full theory
of this complex phenomenon is far from being available, 
we believe that this peculiar behavior requires
a remarkable temperature dependence
of the pairing potential as implied by our ICDW scenario
in the underdoped region and cannot easily be accounted for by
simple models of local pairing.
In the underdoped region  $\kappa^2$ in
Eq. (\ref{fitgamlr}) has a modified temperature dependence 
given by the distance from the critical line $T_{CDW}$ 
Since the pair formation has a  stabilizing effect 
on the ICDW instability, we introduce  $\Delta_{Max}$,
the maximum value of $\Delta({\mbox{\boldmath $k$}})$
in $k$-space as an additional cut-off in Eq. (\ref{fitgamlr})
by assuming
$\kappa^2 \equiv Max \left[c \vert \Delta_{Max}(T)\vert, 
c'(T-T_{CDW}) \right]$. Despite the oversimplified form 
of $\kappa^2$, the $T$ behavior
of $\Delta_{Max}$ bears a striking resemblance with
the analogous quantity recently measured with ARPES in underdoped
Bi-2212 samples \cite{harris}, with a long tail extending up to
large temperatures $T\approx T_{CDW} \approx T^*$ \cite{evora}.

Of course the above BCS treatment only deals with the
amplitude of the gap and says nothing on the
way a true superconducting phase coherence 
is established below a critical temperature $T_c<T^*$. 
For this we have to invoke phase fluctuations as in the
usual (large) negative-U Hubbard model. Here the additional
features (the peaked $q$-dependence of the pairing potential,
and precursors of stripe formation) are expected to
enhance the role of phase fluctuations.

It is also important to emphasize once again that the occurrence of
local pairing prevents the actual establishing of the
ICDW long-range order \cite{notaCCCDGR}, so that $T_{CDW}$ 
and the ICDW-QCP loose
their meaning of a true transition , and they merely
indicate the crossover regions where pairing and strong
 ICDW (dynamical) order selfconsistently interplay.

%
{\it Acknowledgments:} 
Part of this work was carried out with the financial support
of the Istituto Nazionale di Fisica della Materia, Progetto
di Ricerca Avanzata 1996.

%
%

\end{multicols}

\begin{references}
\bibitem{notadisorder} However, disorder can play an important 
role in modifying
$T_{CDW}$ and in determining the $\log T$ behavior of the resistivity
in the insulating CDW phase.
\bibitem {notaBCS} It should, however, be kept in mind that 
even in the uniform (i.e. non-ordered CDW), the evaluation
of a quantitatively reliable superconducting critical temperature
is a difficult task, since the pairing is mediated by singular
interactions.
\bibitem{notaCCCDGR} 
Indeed it was shown in a Kondo-lattice-like
model that superconductivity stabilizes phase separation already
at mean field level\cite{CCCDGR}.
\end{references}
\end{document}